\documentclass[conference]{IEEEtran}
\IEEEoverridecommandlockouts

\usepackage{cite}
\usepackage{amsmath,amssymb,amsfonts}
\usepackage{graphicx}
\usepackage{textcomp}
\usepackage{xcolor}

\usepackage{dcolumn}
\usepackage{bm}
\usepackage{qcircuit}
\usepackage{braket}
\usepackage{float}
\usepackage{algorithm}
\usepackage{algorithmicx}
\usepackage{algpseudocode}
\usepackage{mathtools}
\usepackage{hyperref}
\usepackage{multirow}

\usepackage{url}
\usepackage{subcaption}
\usepackage{makecell}

\def\BibTeX{{\rm B\kern-.05em{\sc i\kern-.025em b}\kern-.08em
    T\kern-.1667em\lower.7ex\hbox{E}\kern-.125emX}}

\begin{document}

\title{Federated Quantum-Train with Batched \\ Parameter Generation \thanks{The views expressed in this article are those of the authors and do not represent the views of Wells Fargo. This article is for informational purposes only. Nothing contained in this article should be construed as investment advice. Wells Fargo makes no express or implied warranties and expressly disclaims all legal, tax, and accounting implications related to this article.}}

\author{
\IEEEauthorblockN{
    Chen-Yu Liu \IEEEauthorrefmark{2}\IEEEauthorrefmark{6}, 
    Samuel Yen-Chi Chen\IEEEauthorrefmark{3}\IEEEauthorrefmark{7}
}

\IEEEauthorblockA{\IEEEauthorrefmark{2}Graduate Institute of Applied Physics, National Taiwan University, Taipei, Taiwan}
\IEEEauthorblockA{\IEEEauthorrefmark{3}Wells Fargo, New York, NY, USA}

\IEEEauthorblockA{Email:
\IEEEauthorrefmark{6}
d10245003@g.ntu.edu.tw, 
\IEEEauthorrefmark{7}
yen-chi.chen@wellsfargo.com
}}

\maketitle

\begin{abstract}
In this work, we introduce the Federated Quantum-Train (QT) framework, which integrates the QT model into federated learning to leverage quantum computing for distributed learning systems. Quantum client nodes employ Quantum Neural Networks (QNNs) and a mapping model to generate local target model parameters, which are updated and aggregated at a central node. Testing with a VGG-like convolutional neural network on the CIFAR-10 dataset, our approach significantly reduces qubit usage from 19 to as low as 8 qubits while reducing generalization error. The QT method mitigates overfitting observed in classical models, aligning training and testing accuracy and improving performance in highly compressed models. Notably, the Federated QT framework does not require a quantum computer during inference, enhancing practicality given current quantum hardware limitations. This work highlights the potential of integrating quantum techniques into federated learning, paving the way for advancements in quantum machine learning and distributed learning systems.
\end{abstract}


\section{Introduction}
Quantum computing (QC) promises potential computational advantages for certain tasks over classical computers, particularly in areas like machine learning (ML) and combinatorial optimization problems \cite{abbas2021power, liu2024quantumlocal, raymond2023hybrid, liu2023hybrid, booth2017partitioning, liu2024parallel, phillipson2021portfolio, liu2022implementation}. Meanwhile, the advances in classical ML and artificial intelligence (AI) have demonstrated amazing capabilities in various tasks \cite{carrasquilla2017machine, seif2019machine, liu2023reinforcement, liu2021random}. With the progress in quantum hardware, it is natural to consider the combination of these two fascinating technologies.
While existing quantum computing devices still suffer from noises and imperfections, a hybrid quantum-classical computing paradigm \cite{bharti2022noisy} which divides computational tasks among quantum and classical computing resources according to their properties to leverage the best part from the both world is proposed. 
Variational quantum algorithms (VQAs) \cite{bharti2022noisy} are the fundamental algorithms framework under this hybrid paradigm. Leading quantum machine learning (QML) methods largely rely on these variational algorithms.
Variational quantum circuits (VQCs) are the building blocks of existing QML models \cite{mitarai2018quantum}. It has been shown theoretically that VQC can outperform classical models when certain conditions are met \cite{abbas2021power, du2020expressive,caro2022generalization}. VQC-based QML models have been shown to be successful in various ML tasks ranging from classification \cite{mitarai2018quantum,chen2022quantumCNN,chen2021end,liu2024quantum,liu2024training,chen2024quantum, chen2023quantum}, time-series modeling \cite{chen2022quantumLSTM, lin2024quantum}, audio and language processing \cite{di2022dawn,stein2023applying,li2023pqlm}, quantum algorithm reconstruction \cite{liu2023practical, liu2023learning},  and reinforcement learning \cite{chen2020variational,lockwood2020reinforcement,chen2022variational,yun2023quantum, liu2024qtrl}.

The great success of modern AI/ML techniques not only depend on good model architecture design but also on the volume of high-quality data and QML is no exception. The requirements of data also raise the privacy concerns in the QML research and application. Among various methods to mitigate the privacy concerns, \emph{federated learning} (FL) is a method in which various participating parties share the locally trained models but not the actual training data to avoid data leakage.

Several FL methods have been proposed in the realm of QML to enhance the privacy-preserving features \cite{chen2021federated,chehimi2022quantum,kwak2023quantum,chehimi2023foundations,kim2023quantum,rofougaran2024federated}. 

While effective, these quantum FL (QFL) methods require the trained models to be used on quantum devices in the inference stage. It poses certain challenges at the moment as there are limited real quantum resources available and it is unclear whether the proposed methods are realistic in the real-world scenarios.

In this paper, we propose a Quantum-Train (QT)-based \cite{liu2024training, lin2024quantum, liu2024quantum} QFL method in which the quantum neural networks (QNN) are trained to generate the well-performing classical neural network weights in the federated setting. Once the training is finished, the QNN is not used during the inference phase.
Our main contributions are: 
\begin{itemize}
    \item \textit{Addressing data encoding issue in QFL}: The QT approach integrated with FL simplifies data handling by using classical inputs and outputs, avoiding the complexities and potential information loss of encoding large datasets into quantum states. This method retains quantum computational advantages without the scaling difficulties of quantum data encoding.

    \item \textit{Reduction of qubit count in QT}: Utilizing the batched parameter generation approach, we reduce the qubit usage from $\lceil \log_2 m \rceil$ to $\lceil \log_2 \lceil \frac{m}{n_{mlp}} \rceil \rceil$, compared to the original QT proposal. Here, $m$ is the number of parameters of the target classical model and $n_{mlp}$ is the batch size in the parameter generation approach. In the example examined in this study, qubit usage is reduced from 19 to as low as 8 qubits.

    \item \textit{Inference without quantum hardware}: The training results of QT are designed to operate seamlessly on classical hardware, eliminating the need for quantum computing resources, unlike conventional QML and QFL. This feature enhances its applicability, especially given the current limited access to quantum computers compared to classical counterparts.
    
\end{itemize}
\section{Federated Learning}
Federated Learning (FL) \cite{mcmahan2017communication} has emerged in response to growing privacy concerns associated with large-scale datasets and cloud-based deep learning \cite{shokri2015privacy}. In the FL framework, the primary components are a \emph{central node} and multiple \emph{client nodes}. The central node maintains the \emph{global model} and collects trained parameters from the client nodes. It then performs an \emph{aggregation} process to update the global model, which is subsequently shared with all client nodes. The client nodes locally train the received model using their own data, which typically constitutes a small subset of the overall dataset.
The concept of FL has been explored in the field of QML since the publications \cite{chen2021federated, chehimi2022quantum}. In \cite{chen2021federated}, the authors examined the simplest form of QFL utilizing hybrid quantum-classical models. In this approach, a pre-trained CNN compresses input images into a dimension manageable by a VQC. The locally trained hybrid model parameters are then uploaded to a central server, which aggregates these parameters and distributes the updated model to all participants. This framework has been further enhanced to process sequential data using a federated quantum LSTM network \cite{Chehimi2024FedQLSTM}. The study \cite{chehimi2022quantum} delves into a more advanced scenario where QFL processes quantum states instead of classical images.
While QFL can mitigate the risk of direct leakage of training datasets, it remains vulnerable to attacks that can extract training data entries from the trained models themselves. Such attacks pose a significant threat to data privacy. To address this issue, \cite{rofougaran2024federated} explores the integration of differentially-private gradient optimizers with QFL, aiming to enhance the privacy of QML models.
QFL can be further extended to scenarios where training is conducted on encrypted data, as demonstrated in \cite{chu2023cryptoqfl}.
QFL can be applied in diverse scenarios, including autonomous vehicles \cite{kim2023quantum} and quantum fuzzy learning \cite{qu2024quantum}.
\section{Variational Quantum Circuits and Quantum-Train}
\begin{figure*}[ht]
\centering
\includegraphics[scale=0.21]{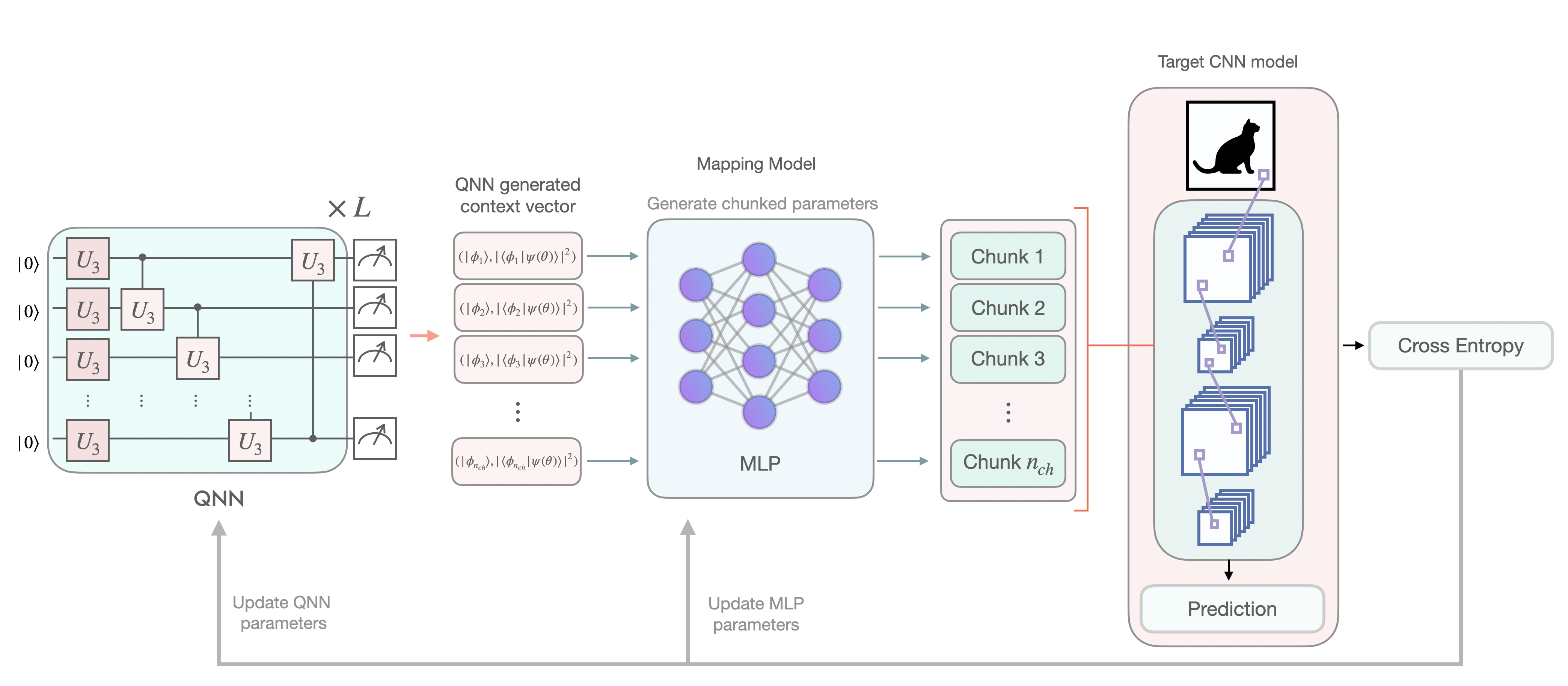}
\caption{Illustration of the QT method with batched parameter generation. The target model parameters $\omega$ are divided into chunks, each containing $n_{mlp}$ parameters. The QNN generates measurement probabilities which are mapped by $\Tilde{G}_{\beta}$ to produce a batch of parameters $\vec{\omega}_i$. This approach reduces qubit usage and maintains the benefits of the QT method.
}
\label{fig:QuantumTrain}
\end{figure*}

At the core of the QML scheme, VQCs play a pivotal role by providing the parameterized ansatz that forms the function approximator for learning tasks. A typical VQC used as a QNN is depicted on the left side of Fig.~\ref{fig:QuantumTrain}. The process begins with the initial state $|0 \rangle^{\otimes N}$, where $N$ is the number of qubits. This is followed by parameterized single-qubit and two-qubit unitary operations $U_3$ and controlled-$U_3$ ($CU_3$) gates, characterized by their matrix representations:
\begin{eqnarray}
    && U_3(\mu, \varphi, \lambda) = \left[ \begin{array}{cc}
    \cos(\mu/2) & -e^{i \lambda} \sin(\mu/2) \\
    e^{i \varphi} \sin(\mu/2) & e^{i(\varphi + \lambda)} \cos(\mu/2)
    \end{array} \right], \\
    &&  CU_3 = I \otimes |0\rangle \langle 0 | + U_3(\mu, \varphi, \lambda) \otimes |1\rangle \langle 1 |,
\end{eqnarray}
The parameterized quantum state (QNN) can then be described as:
\begin{equation}
\label{vqc}
|\psi(\theta) \rangle = \left(\prod_i CU_3^{i, i+1} \prod_j U_3^j \right)^L |0\rangle^{\otimes N},
\end{equation}
where $i$ and $j$ are qubit indices, and $L$ is the number of repetitions. The proposed vanilla QT \cite{liu2024quantum} is as follows: consider a target neural network model with parameters $\omega$, where $\omega = (\omega_1, \omega_2, \ldots, \omega_m)$ and $m$ is the total number of parameters. Instead of updating all $m$ parameters as in conventional ML, QT utilizes $ |\psi (\theta) \rangle $, a QNN with $N = \lceil \log_2 m \rceil$ qubits, to generate $2^N$ distinct measurement probabilities $ |\langle \phi_i | \psi (\theta) \rangle|^2 $ for $ i \in \{1, 2, \ldots, 2^N\} $, where $ |\phi_i \rangle $ is the $i$-th basis state. These probabilities are then input into a mapping model $ G_{\beta} $, a multi-layer perceptron (MLP) type classical neural network with parameters $\beta$. 
The first $m$ basis measurement result probabilities, along with the vector representations of the corresponding basis states $|\phi_i \rangle$, are mapped from values bounded between 0 and 1 to $-\infty$ and $\infty$ using the following equation:
\begin{equation}
    G_{\beta} (|\phi_i \rangle, |\langle \phi_i | \psi (\theta) \rangle|^2) = \omega_i, \quad i = 1, 2, \ldots, m.
\end{equation}
Here, it can be observed that the parameter $\omega$ of the target model is generated from the QNN $|\psi(\theta) \rangle$ and the mapping model $G_{\beta}$. Notably, the required number of parameters for both $\theta$ and $\beta$ scales as $O(\text{polylog}(m))$ \cite{liu2024quantum}, allowing for the effective training of the target model with $m$ parameters by only tuning $O(\text{polylog}(m))$ parameters of $\theta$ and $\beta$. Unlike conventional QML approaches, which require the QNN during the inference stage, the QT approach decouples the quantum computing resource after training. Since the QNN is used solely for generating the parameters of the target model, the resulting trained model is a classical neural network. This classical model can then be executed entirely on classical computing hardware. This characteristic is particularly practical given that quantum computing hardware is currently a relatively rare and expensive resource.

\begin{figure*}[ht]
\centering
\includegraphics[scale=0.19]{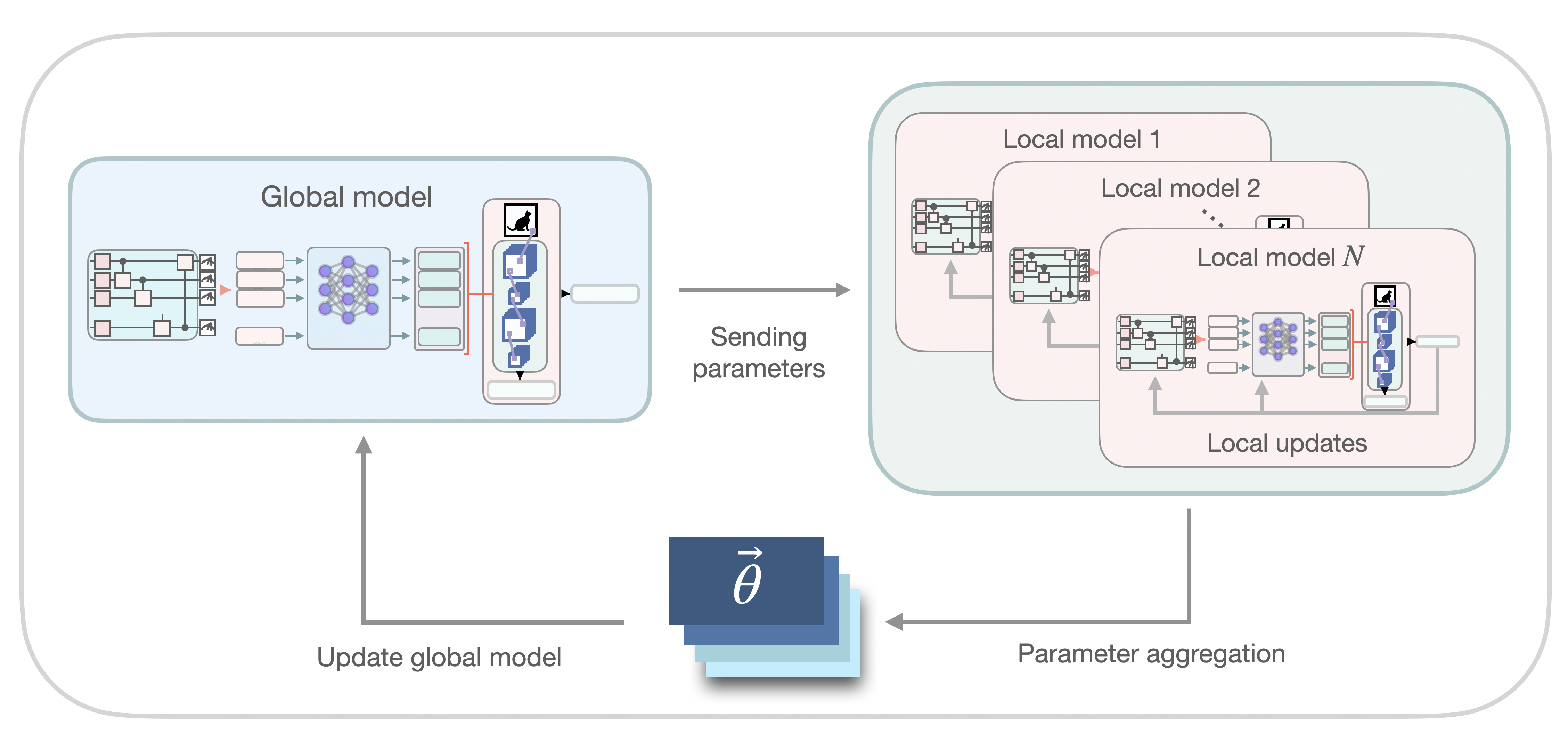}
\caption{Diagram of the Federated QT framework. Each quantum client node employs a QNN and a mapping model $\Tilde{G}_{\beta}$ to generate local target model parameters. These parameters are updated based on local datasets and aggregated at a central node to update the global model. The framework leverages quantum computing for training while only requiring classical computation during inference.
}
\label{fig:FederatedLearning}
\end{figure*}

\section{Quantum-Train with batched parameter generation}

Building upon the previously proposed QT method, which generates a single parameter of the target network model from a single basis measurement probability, this study introduces a batch parameter generation approach. This method generates a batch of parameters from a single basis measurement probability, as illustrated in Fig.~\ref{fig:QuantumTrain}. In this approach, the $m$ parameters of the target model are divided into $n_{ch}$ chunks, each containing $n_{mlp}$ parameters, such that $n_{ch} = \lceil m / n_{mlp} \rceil$. The mapping model, now denoted as $\Tilde{G}_{\beta}$, takes as input $|\phi_i \rangle$ and $|\langle \phi_i | \psi (\theta) \rangle|^2$ and generates a batch of parameters in $\omega$ of size $n_{mlp}$:
\begin{eqnarray}
    && \Tilde{G}_{\beta} (|\phi_i \rangle, |\langle \phi_i | \psi (\theta) \rangle|^2) = \vec{\omega}_i, \quad i = 1, 2, \ldots, n_{ch}, \\
    && \vec{\omega}_i = ( \omega_{i,1}, \omega_{i,2}, ... \omega_{i,j} ), \quad j = 1,2, \ldots, n_{mlp}. 
\end{eqnarray}
This setup is realized through a decoder-like architecture of the MLP in the mapping model $\Tilde{G}_{\beta}$, where the output size is expanded from 1 to $n_{mlp}$, or \textsf{mlp\_out} in the following syntax. Consequently, the qubit usage $N$ is reduced from $N = \lceil \log_2 m \rceil$ to 
\begin{equation}
\label{eq:QTBG_qubits_usage}
    N = \lceil \log_2 n_{ch} \rceil = \lceil \log_2 \lceil \frac{m}{n_{mlp}} \rceil \rceil,
\end{equation}
effectively saving approximately $\lceil \log_2 n_{mlp} \rceil$ qubits from the original QT proposal, the original method can be considered as a special case with $n_{mlp} = 1$. Reducing qubit usage also mitigates the issue of the exponential requirement of measurement shots, as mentioned in the original QT study. The remaining training process is similar to the vanilla QT method, as depicted in Fig.~\ref{fig:QuantumTrain}.

\section{Federated Quantum-Train}
Following the original idea of FL and QFL, we introduce the concept of the QT model within the federated framework. In this approach, each quantum client node employs QNN $|\psi (\theta) \rangle$ and mapping model $\Tilde{G}_{\beta}$ to generate the local target model parameters. During each training round, every quantum client nodes update their QNN parameters and the associated mapping model parameters based on their local datasets. These updated parameters are sent to the central node, where they are aggregated to update the global model, as depicted in Fig.~\ref{fig:FederatedLearning}. This process ensures that the global model benefits from the QT performed at each client node, leading to improved performance and efficiency. By integrating the QT model into FL, we leverage the advantages of quantum computing to reduce the number of training parameters and enhance the scalability of distributed learning systems. Notably, compared to traditional QFL, federated QT does not require a quantum computer during the inference stage. 

\section{Result and Discussion}

\begin{figure}[ht]
\centering
\includegraphics[scale=0.22]{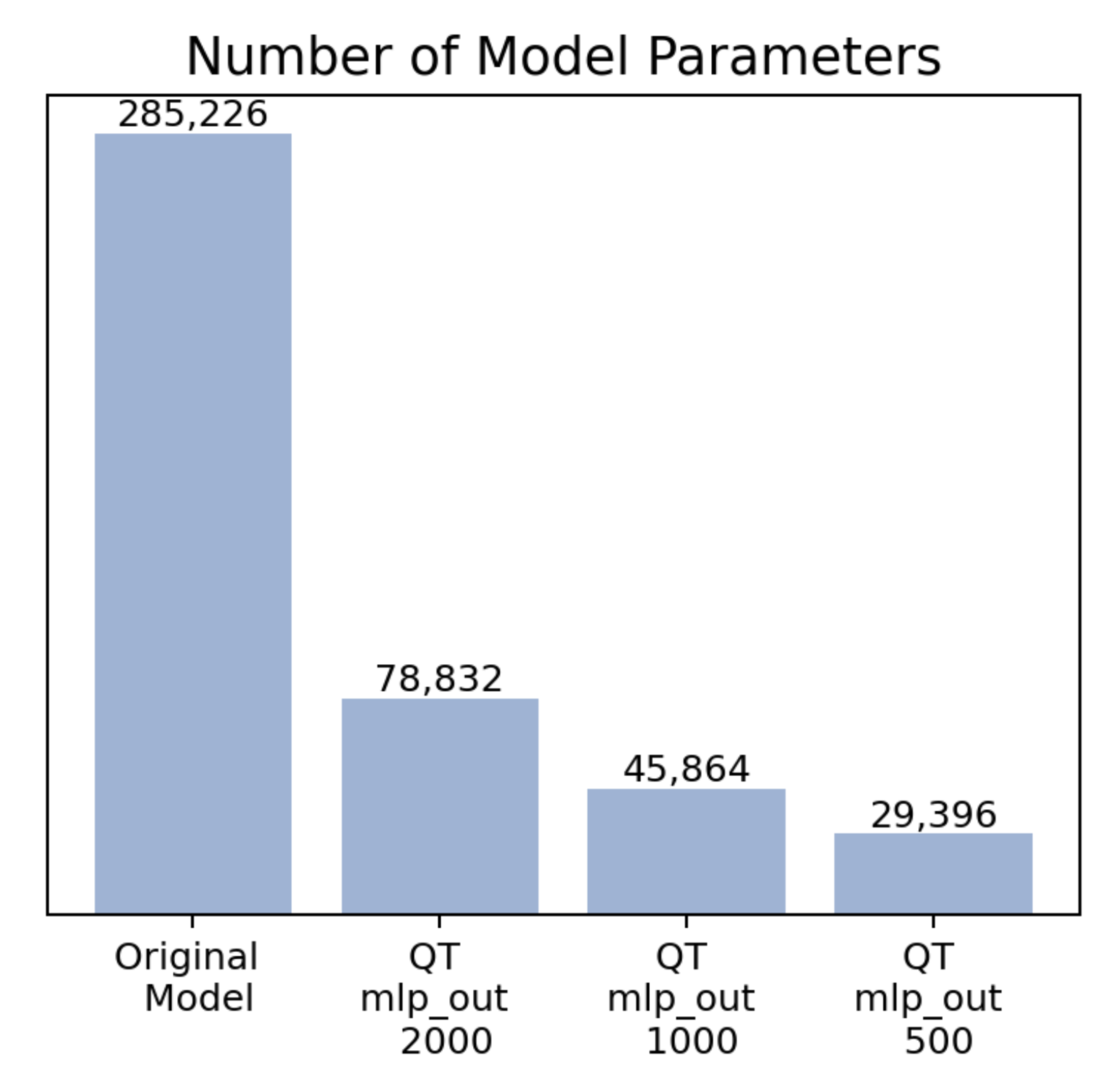}
\caption{Comparison of the number of training parameters for different models used in this study.}
\label{fig:ParameterCount}
\end{figure}

To examine the applicability of the proposed federated QT framework, we tested it using a VGG-like convolutional neural network (CNN) structure with the CIFAR-10 dataset. The target CNN model has 285226 parameters. We tested three different QT setups: $n_{mlp} \in \{2000, 1000, 500\}$, while fixing the repetition $L = 5$. The required qubit usage, derived from Eq.~\ref{eq:QTBG_qubits_usage}, is 8, 9, and 10 qubits, respectively. Compared to the original QT with the same CNN model \cite{liu2024quantum}, which required 19 qubits, our new batched parameter generation significantly reduces the qubit usage. Fig.~\ref{fig:ParameterCount} illustrates the number of model parameters for the models investigated in this study, with $n_{mlp}$ denoted as \textsf{mlp\_out}.

\begin{figure*}[ht]
\centering
\includegraphics[scale=0.22]{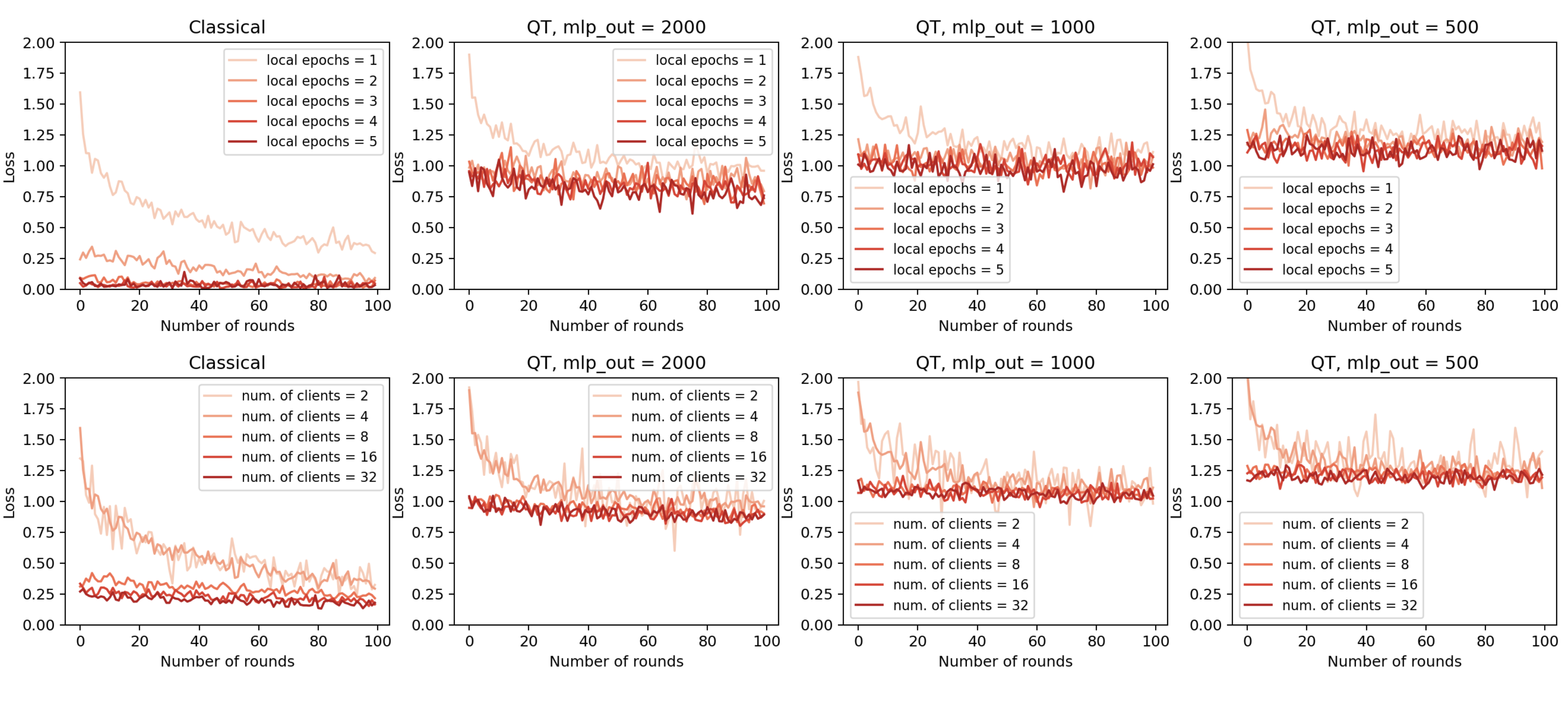}
\caption{The Cross-Entropy loss for the CIFAR-10 image classification task across multiple training rounds. The upper row illustrates the effect of varying the number of local epochs and \textsf{mlp\_out}, with the number of clients fixed at 4. The lower row demonstrates the effect of varying the number of clients while keeping the local epoch fixed at 1. The results indicate that a larger number of local epochs and clients lead to lower loss values, suggesting improved model performance.
}
\label{fig:ResultLoss}
\end{figure*}

In the upper row of Fig.~\ref{fig:ResultLoss}, the Cross-Entropy loss for the CIFAR-10 image classification task, involving 10 classes over multiple rounds, is presented with different setups of local epochs and \textsf{mlp\_out}, while fixing the number of clients at 4. It can be observed that a larger number of local epochs leads to lower loss values. This outcome is expected, as the model undergoes more frequent updates, providing more opportunities to correct incorrect predictions.

In the lower row of Fig.~\ref{fig:ResultLoss}, the local epoch is fixed to 1, and the effect of varying the number of clients is investigated. In this investigation, the dataset is divided into as many pieces as the number of clients. Interestingly, increasing the number of clients results in better performance in terms of training loss. This improvement can be attributed to the flexibility provided by different local models, which adjust distinct sets of parameters corresponding to different parts of the dataset. The parameter aggregation step then incorporates these updates from diverse perspectives, enhancing the overall model performance.

\begin{figure*}[ht]
\centering
\includegraphics[scale=0.225]{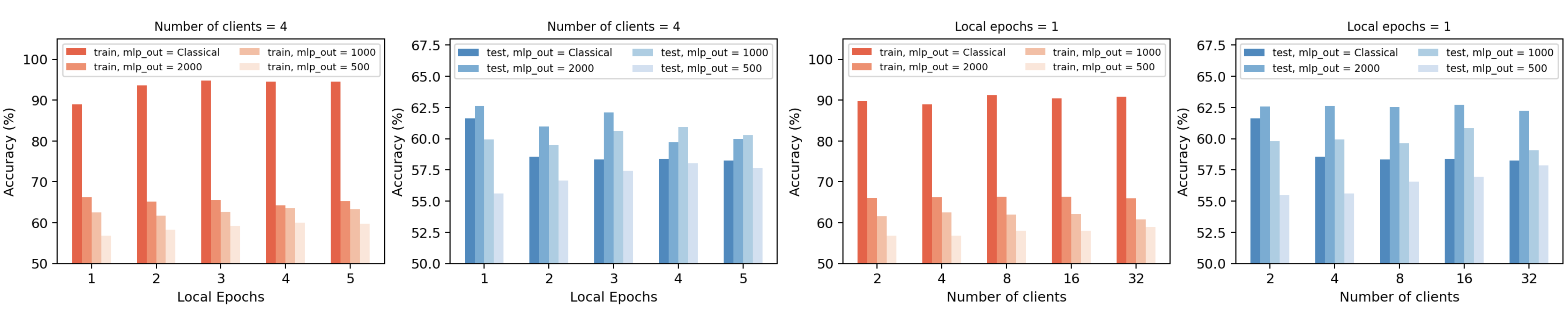}
\caption{The training and testing accuracy for the models investigated in Fig.~\ref{fig:ResultLoss}. The results reveal significant overfitting in the purely classical model, with high training accuracy but lower testing accuracy compared to the $n_{mlp} = 2000$ case. The QT method reduces the deviation between training and testing accuracy, highlighting its advantage in minimizing generalization error. Additionally, the federated QT framework demonstrates improved performance in highly compressed models, such as the $n_{mlp} = 500$ case, which uses only about $10\%$ of the original CNN model’s parameters.}
\label{fig:ResultAcc}
\end{figure*}

A noticeable trend in the figures is that models with more parameters tend to perform better in terms of training loss. This can be attributed to the increased expressiveness of the corresponding models, which enhances their ability to fit the training data. However, this observation only indicates the model’s effectiveness on the training dataset and does not necessarily reflect its general performance on unseen data.

As illustrated in Fig.~\ref{fig:ResultAcc}, we present the testing and training accuracy of the models investigated in Fig.~\ref{fig:ResultLoss}. A notable observation is the significant overfitting in the purely classical case. While the training accuracy of the classical model is extremely high, its testing accuracy is slightly lower than that of the $n_{mlp} = 2000$ case. This behavior underscores an advantage of the QT method, as highlighted in previous studies \cite{liu2024quantum}: the QT method can reduce the deviation between training and testing accuracy, which is proportional to the generalization error. Moreover, our batched parameter generation approach not only significantly reduces qubit usage but also preserves the advantage of generalization error reduction inherent in the vanilla QT method.

While there is no clear trend for different local epochs and the number of clients in the classical, $n_{mlp} = 2000$, and $n_{mlp} = 1000$ cases, the $n_{mlp} = 500$ case shows an increase in both training and testing accuracy with an increase in local epochs and the number of clients. This behavior demonstrates that the federated framework combined with QT can improve models with extreme compression, such as the $n_{mlp} = 500$ case, which uses only about $10\%$ of the original CNN model’s parameters.

\section{Conclusion}
\label{sec:conclusion}
In this work, we introduced the Federated QT framework, integrating the QT model into federated learning to leverage quantum computing for distributed learning systems. Each quantum client node employs QNNs and a mapping model $\Tilde{G}_{\beta}$ to generate local target model parameters. These parameters are updated based on local datasets and aggregated at a central node, enhancing the global model through quantum-enhanced training.

Our experiments, using a VGG-like CNN on the CIFAR-10 dataset, demonstrate the efficacy of the Federated QT framework. We tested three different QT setups with varying $n_{mlp}$ values, significantly reducing the required qubit usage compared to the original QT method. Specifically, our batched parameter generation approach reduced qubit usage from 19 to as low as 8, while maintaining the benefits of generalization error reduction.

Results indicate that models with more parameters perform better in training loss due to increased expressiveness, but overfitting was observed in purely classical models. The QT method mitigated this issue, resulting in a closer alignment between training and testing accuracy. The federated framework combined with QT also showed improved performance in highly compressed models, such as the $n_{mlp} = 500$ case, which uses only about $10\%$ of the original CNN model’s parameters.

The Federated QT framework provides a scalable and efficient approach to distributed learning, utilizing quantum computing to reduce training parameters and enhance model performance. Notably, QT does not require a quantum computer during the inference stage, making it highly practical given the current limitations of quantum hardware. Our findings highlight the practical benefits of integrating quantum techniques into federated learning, paving the way for future advancements in QML and distributed learning systems.

\bibliographystyle{IEEEtran}
\bibliography{bib/tools,bib/vqc,bib/qml_examples,bib/quantum_fl, bib/ml_examples, bib/hybrid_co_examples,bib/classical_fl}

\begin{thebibliography}{10}
\providecommand{\url}[1]{#1}
\csname url@samestyle\endcsname
\providecommand{\newblock}{\relax}
\providecommand{\bibinfo}[2]{#2}
\providecommand{\BIBentrySTDinterwordspacing}{\spaceskip=0pt\relax}
\providecommand{\BIBentryALTinterwordstretchfactor}{4}
\providecommand{\BIBentryALTinterwordspacing}{\spaceskip=\fontdimen2\font plus
\BIBentryALTinterwordstretchfactor\fontdimen3\font minus \fontdimen4\font\relax}
\providecommand{\BIBforeignlanguage}[2]{{%
\expandafter\ifx\csname l@#1\endcsname\relax
\typeout{** WARNING: IEEEtran.bst: No hyphenation pattern has been}%
\typeout{** loaded for the language `#1'. Using the pattern for}%
\typeout{** the default language instead.}%
\else
\language=\csname l@#1\endcsname
\fi
#2}}
\providecommand{\BIBdecl}{\relax}
\BIBdecl

\bibitem{abbas2021power}
A.~Abbas, D.~Sutter, C.~Zoufal, A.~Lucchi, A.~Figalli, and S.~Woerner, ``The power of quantum neural networks,'' \emph{Nature Computational Science}, vol.~1, no.~6, pp. 403--409, 2021.

\bibitem{liu2024quantumlocal}
C.-Y. Liu, H.~Matsuyama, W.-h. Huang, and Y.~Yamashiro, ``Quantum local search for traveling salesman problem with path-slicing strategy,'' \emph{arXiv preprint arXiv:2407.13616}, 2024.

\bibitem{raymond2023hybrid}
J.~Raymond, R.~Stevanovic, W.~Bernoudy, K.~Boothby, C.~C. McGeoch, A.~J. Berkley, P.~Farr{\'e}, J.~Pasvolsky, and A.~D. King, ``Hybrid quantum annealing for larger-than-qpu lattice-structured problems,'' \emph{ACM Transactions on Quantum Computing}, vol.~4, no.~3, pp. 1--30, 2023.

\bibitem{liu2023hybrid}
C.-Y. Liu and H.-S. Goan, ``Hybrid gate-based and annealing quantum computing for large-size ising problems,'' \emph{Bulletin of the American Physical Society}, vol.~68, 2023.

\bibitem{booth2017partitioning}
M.~Booth, S.~P. Reinhardt, and A.~Roy, ``Partitioning optimization problems for hybrid classical,'' \emph{quantum execution. Technical Report}, pp. 01--09, 2017.

\bibitem{liu2024parallel}
C.-Y. Liu and K.-C. Chen, ``Parallel quantum local search via evolutionary mechanism,'' \emph{arXiv preprint arXiv:2406.06445}, 2024.

\bibitem{phillipson2021portfolio}
F.~Phillipson and H.~S. Bhatia, ``Portfolio optimisation using the d-wave quantum annealer,'' in \emph{International Conference on Computational Science}.\hskip 1em plus 0.5em minus 0.4em\relax Springer, 2021, pp. 45--59.

\bibitem{liu2022implementation}
C.-Y. Liu, H.-Y. Wang, P.-Y. Liao, C.-J. Lai, and M.-H. Hsieh, ``Implementation of trained factorization machine recommendation system on quantum annealer,'' \emph{arXiv preprint arXiv:2210.12953}, 2022.

\bibitem{carrasquilla2017machine}
J.~Carrasquilla and R.~G. Melko, ``Machine learning phases of matter,'' \emph{Nature Physics}, vol.~13, no.~5, pp. 431--434, 2017.

\bibitem{seif2019machine}
A.~Seif, M.~Hafezi, and C.~Jarzynski, ``Machine learning the thermodynamic arrow of time,'' \emph{Nature Physics}, vol.~17, no.~1, pp. 105--113, 2021.

\bibitem{liu2023reinforcement}
C.-Y. Liu and H.-S. Goan, ``Reinforcement learning quantum local search,'' in \emph{2023 IEEE International Conference on Quantum Computing and Engineering (QCE)}, vol.~2.\hskip 1em plus 0.5em minus 0.4em\relax IEEE, 2023, pp. 246--247.

\bibitem{liu2021random}
C.-Y. Liu and D.-W. Wang, ``Random sampling neural network for quantum many-body problems,'' \emph{Physical Review B}, vol. 103, no.~20, p. 205107, 2021.

\bibitem{bharti2022noisy}
K.~Bharti, A.~Cervera-Lierta, T.~H. Kyaw, T.~Haug, S.~Alperin-Lea, A.~Anand, M.~Degroote, H.~Heimonen, J.~S. Kottmann, T.~Menke \emph{et~al.}, ``Noisy intermediate-scale quantum algorithms,'' \emph{Reviews of Modern Physics}, vol.~94, no.~1, p. 015004, 2022.

\bibitem{mitarai2018quantum}
K.~Mitarai, M.~Negoro, M.~Kitagawa, and K.~Fujii, ``Quantum circuit learning,'' \emph{Physical Review A}, vol.~98, no.~3, p. 032309, 2018.

\bibitem{du2020expressive}
Y.~Du, M.-H. Hsieh, T.~Liu, and D.~Tao, ``Expressive power of parametrized quantum circuits,'' \emph{Physical Review Research}, vol.~2, no.~3, p. 033125, 2020.

\bibitem{caro2022generalization}
M.~C. Caro, H.-Y. Huang, M.~Cerezo, K.~Sharma, A.~Sornborger, L.~Cincio, and P.~J. Coles, ``Generalization in quantum machine learning from few training data,'' \emph{Nature communications}, vol.~13, no.~1, pp. 1--11, 2022.

\bibitem{chen2022quantumCNN}
S.~Y.-C. Chen, T.-C. Wei, C.~Zhang, H.~Yu, and S.~Yoo, ``Quantum convolutional neural networks for high energy physics data analysis,'' \emph{Physical Review Research}, vol.~4, no.~1, p. 013231, 2022.

\bibitem{chen2021end}
S.~Y.-C. Chen, C.-M. Huang, C.-W. Hsing, and Y.-J. Kao, ``An end-to-end trainable hybrid classical-quantum classifier,'' \emph{Machine Learning: Science and Technology}, vol.~2, no.~4, p. 045021, 2021.

\bibitem{liu2024quantum}
C.-Y. Liu, E.-J. Kuo, C.-H.~A. Lin, J.~G. Young, Y.-J. Chang, M.-H. Hsieh, and H.-S. Goan, ``Quantum-train: Rethinking hybrid quantum-classical machine learning in the model compression perspective,'' \emph{arXiv preprint arXiv:2405.11304}, 2024.

\bibitem{liu2024training}
C.-Y. Liu, E.-J. Kuo, C.-H.~A. Lin, S.~Chen, J.~G. Young, Y.-J. Chang, and M.-H. Hsieh, ``Training classical neural networks by quantum machine learning,'' \emph{arXiv preprint arXiv:2402.16465}, 2024.

\bibitem{chen2024quantum}
K.-C. Chen, X.~Li, X.~Xu, Y.-Y. Wang, and C.-Y. Liu, ``Quantum-hpc framework with multi-gpu-enabled hybrid quantum-classical workflow: Applications in quantum simulations,'' \emph{arXiv preprint arXiv:2403.05828}, 2024.

\bibitem{chen2023quantum}
K.-C. Chen, X.~Xu, H.~Makhanov, H.-H. Chung, and C.-Y. Liu, ``Quantum-enhanced support vector machine for large-scale stellar classification with gpu acceleration,'' \emph{arXiv preprint arXiv:2311.12328}, 2023.

\bibitem{chen2022quantumLSTM}
S.~Y.-C. Chen, S.~Yoo, and Y.-L.~L. Fang, ``Quantum long short-term memory,'' in \emph{ICASSP 2022-2022 IEEE International Conference on Acoustics, Speech and Signal Processing (ICASSP)}.\hskip 1em plus 0.5em minus 0.4em\relax IEEE, 2022, pp. 8622--8626.

\bibitem{lin2024quantum}
C.-H.~A. Lin, C.-Y. Liu, and K.-C. Chen, ``Quantum-train long short-term memory: Application on flood prediction problem,'' \emph{arXiv preprint arXiv:2407.08617}, 2024.

\bibitem{di2022dawn}
R.~Di~Sipio, J.-H. Huang, S.~Y.-C. Chen, S.~Mangini, and M.~Worring, ``The dawn of quantum natural language processing,'' in \emph{ICASSP 2022-2022 IEEE International Conference on Acoustics, Speech and Signal Processing (ICASSP)}.\hskip 1em plus 0.5em minus 0.4em\relax IEEE, 2022, pp. 8612--8616.

\bibitem{stein2023applying}
J.~Stein, I.~Christ, N.~Kraus, M.~B. Mansky, R.~M{\"u}ller, and C.~Linnhoff-Popien, ``Applying qnlp to sentiment analysis in finance,'' in \emph{2023 IEEE International Conference on Quantum Computing and Engineering (QCE)}, vol.~2.\hskip 1em plus 0.5em minus 0.4em\relax IEEE, 2023, pp. 20--25.

\bibitem{li2023pqlm}
S.~S. Li, X.~Zhang, S.~Zhou, H.~Shu, R.~Liang, H.~Liu, and L.~P. Garcia, ``Pqlm-multilingual decentralized portable quantum language model,'' in \emph{ICASSP 2023-2023 IEEE International Conference on Acoustics, Speech and Signal Processing (ICASSP)}.\hskip 1em plus 0.5em minus 0.4em\relax IEEE, 2023, pp. 1--5.

\bibitem{liu2023practical}
C.-Y. Liu, ``Practical quantum search by variational quantum eigensolver on noisy intermediate-scale quantum hardware,'' in \emph{2023 International Conference on Computational Science and Computational Intelligence (CSCI)}.\hskip 1em plus 0.5em minus 0.4em\relax IEEE, 2023, pp. 397--403.

\bibitem{liu2023learning}
C.-Y. Liu, C.-H.~A. Lin, and K.-C. Chen, ``Learning quantum phase estimation by variational quantum circuits,'' \emph{arXiv preprint arXiv:2311.04690}, 2023.

\bibitem{chen2020variational}
S.~Y.-C. Chen, C.-H.~H. Yang, J.~Qi, P.-Y. Chen, X.~Ma, and H.-S. Goan, ``Variational quantum circuits for deep reinforcement learning,'' \emph{IEEE access}, vol.~8, pp. 141\,007--141\,024, 2020.

\bibitem{lockwood2020reinforcement}
O.~Lockwood and M.~Si, ``Reinforcement learning with quantum variational circuit,'' in \emph{Proceedings of the AAAI conference on artificial intelligence and interactive digital entertainment}, vol.~16, no.~1, 2020, pp. 245--251.

\bibitem{chen2022variational}
S.~Y.-C. Chen, C.-M. Huang, C.-W. Hsing, H.-S. Goan, and Y.-J. Kao, ``Variational quantum reinforcement learning via evolutionary optimization,'' \emph{Machine Learning: Science and Technology}, vol.~3, no.~1, p. 015025, 2022.

\bibitem{yun2023quantum}
W.~J. Yun, J.~Park, and J.~Kim, ``Quantum multi-agent meta reinforcement learning,'' in \emph{Proceedings of the AAAI Conference on Artificial Intelligence}, vol.~37, no.~9, 2023, pp. 11\,087--11\,095.

\bibitem{liu2024qtrl}
C.-Y. Liu, C.-H.~A. Lin, C.-H.~H. Yang, K.-C. Chen, and M.-H. Hsieh, ``Qtrl: Toward practical quantum reinforcement learning via quantum-train,'' \emph{arXiv preprint arXiv:2407.06103}, 2024.

\bibitem{chen2021federated}
S.~Y.-C. Chen and S.~Yoo, ``Federated quantum machine learning,'' \emph{Entropy}, vol.~23, no.~4, p. 460, 2021.

\bibitem{chehimi2022quantum}
M.~Chehimi and W.~Saad, ``Quantum federated learning with quantum data,'' in \emph{ICASSP 2022-2022 IEEE International Conference on Acoustics, Speech and Signal Processing (ICASSP)}.\hskip 1em plus 0.5em minus 0.4em\relax IEEE, 2022, pp. 8617--8621.

\bibitem{kwak2023quantum}
Y.~Kwak, W.~J. Yun, J.~P. Kim, H.~Cho, J.~Park, M.~Choi, S.~Jung, and J.~Kim, ``Quantum distributed deep learning architectures: Models, discussions, and applications,'' \emph{ICT Express}, vol.~9, no.~3, pp. 486--491, 2023.

\bibitem{chehimi2023foundations}
M.~Chehimi, S.~Y.-C. Chen, W.~Saad, D.~Towsley, and M.~Debbah, ``Foundations of quantum federated learning over classical and quantum networks,'' \emph{IEEE Network}, 2023.

\bibitem{kim2023quantum}
J.~Kim, ``Quantum federated learning for vehicular computing scenarios,'' in \emph{2023 14th International Conference on Information and Communication Technology Convergence (ICTC)}.\hskip 1em plus 0.5em minus 0.4em\relax IEEE, 2023, pp. 168--172.

\bibitem{rofougaran2024federated}
R.~Rofougaran, S.~Yoo, H.-H. Tseng, and S.~Y.-C. Chen, ``Federated quantum machine learning with differential privacy,'' in \emph{ICASSP 2024-2024 IEEE International Conference on Acoustics, Speech and Signal Processing (ICASSP)}.\hskip 1em plus 0.5em minus 0.4em\relax IEEE, 2024, pp. 9811--9815.

\bibitem{mcmahan2017communication}
B.~McMahan, E.~Moore, D.~Ramage, S.~Hampson, and B.~A. y~Arcas, ``Communication-efficient learning of deep networks from decentralized data,'' in \emph{Artificial Intelligence and Statistics}.\hskip 1em plus 0.5em minus 0.4em\relax PMLR, 2017, pp. 1273--1282.

\bibitem{shokri2015privacy}
R.~Shokri and V.~Shmatikov, ``Privacy-preserving deep learning,'' in \emph{Proceedings of the 22nd ACM SIGSAC conference on computer and communications security}, 2015, pp. 1310--1321.

\bibitem{Chehimi2024FedQLSTM}
\BIBentryALTinterwordspacing
M.~Chehimi, S.~Y.-C. Chen, W.~Saad, and S.~Yoo, ``Federated quantum long short-term memory (fedqlstm),'' \emph{Quantum Machine Intelligence}, vol.~6, no.~2, p.~43, Jul 2024. [Online]. Available: \url{https://doi.org/10.1007/s42484-024-00174-z}
\BIBentrySTDinterwordspacing

\bibitem{chu2023cryptoqfl}
C.~Chu, L.~Jiang, and F.~Chen, ``Cryptoqfl: quantum federated learning on encrypted data,'' in \emph{2023 IEEE International Conference on Quantum Computing and Engineering (QCE)}, vol.~1.\hskip 1em plus 0.5em minus 0.4em\relax IEEE, 2023, pp. 1231--1237.

\bibitem{qu2024quantum}
Z.~Qu, L.~Zhang, and P.~Tiwari, ``Quantum fuzzy federated learning for privacy protection in intelligent information processing,'' \emph{IEEE Transactions on Fuzzy Systems}, 2024.

\end{thebibliography}

\end{document}